\documentstyle[psfig,12pt]{article}
\begin{document}
\title{Evolution of Structure Functions And Their Moments\\
In Chiral Field Theory}
\author{Kakali Ray-Maity\\and\\
Padmanabha Dasgupta\\ 
Department of Physics, University of
Kalyani\\ West Bengal, India 741 235
\date{}}
\maketitle
\vskip 2cm
\begin{center}
{\bf Abstract}
\end{center}
Evolution of structure functions and their moments at low and 
moderate $Q^2$ is studied in the chiral field theory. Evolution equations 
based on perturbation expansion in the coupling constant of the 
effective theory are derived and solved for the moments. The kernels of 
evolution arising from different processes have been calculated with 
contributions from direct and cross channels, the interference terms 
being non-negligible in the kinematic regions under consideration. This 
is shown to lead to flavor-dependence of the kernels which manifests in 
observable effects. The invalidity of the probabilistic approach to 
the evolution process is also pointed out.
\vfil\eject
\section{Introduction}

 While perturbative QCD has been reasonably successful in describing
 the scale dependence of the structure functions at asymptotically
 large $Q^2$ [1], phenomena observed experimentally at intermediate
 $Q^2$ seem to demand consideration of nonperturbative effects[2],
 the observed departure of the Gottfried sum from the parton model
 value providing a notable example [3]. Effective theories based on
 $qq\pi$ vertex, which are expected to incorporate the relevant
 nonperturbative effects, give simple, and at least qualitative,
 explanation of the observed phenomena[4-6].
 
 In such theories, the QCD evolution of the effective parton 
 densities due to quark-gluon interaction is replaced essentially
 by evolution due to quark pion interaction. The evolution equations
 proposed in these theories involve evolution kernels or splitting
 functions which can be computed from the virtual photon cross sections
for pion emission processes. While in perturbative QCD, the evolution
kernels result directly from the Weisz\"acker-Williams procedure
and admit simple probabilistic interpretation, we point out in this
paper that the correct evolution kernels in pion-emission theories
are not consistent with such interpretation.
  
Although we illustrate the point with the specific example
of the chiral field theory [7] used in [4], the conclusions are
absolutely general.

 The effective quark densities are defined in the Altarelli
-Parisi formalism through the use of leading log approximation.
In effective theories, the lack of asymptotoic freedom requires
a different procedure for defining the effective quark densities.
We discuss this in the next section. In section 3, the evolution
equations for the effective parton densities thus defined are
obtained. In section 4, we also show the evolution equations
that would follow from the probabilistic approach in a
kinematic region where such approach may be supposed to be
valid and point out the difficulties that would result from
such a procedure. In section 5, we discuss the evolution
of the moments of the structure function that follow
from the evolution equations derived in section 3. 
 
\section{Effective quark densities}
\par The standard theoretical approach to the $Q^2$-dependence
of the structure functions begins with the formula
\begin{equation}
F_2^{eN}(x,Q^2)/x=\sum_ie_i^2\int{dy\over y}q_{i\circ}(y)
\sigma^{\gamma^*q_i}(x/y,Q^2)
\end{equation}
where $q_{i\circ}$ is the naive parton model density of the $i$-th
parton in the target nucleon and $\sigma^{\gamma^*q_i}$ is the ${
\gamma^*q_i}$ cross section integrated over $p_T^2$ and made 
adimensional by taking out a scale factor [1]. Correction to the naive parton
model results comes entirely from the cross section which in QCD and in
effective theories is quite different from the $Q^2$-independent point-like
cross section of the former.
\par To analyze the scale dependence of the structure functions from
(2.1), a convenient formalism is obtained by introducing the effective
parton densities $q_i(x,Q^2)$ and setting up evolution equations for
them. It is well known that in perturbative QCD this is done in the
leading logarithm approximation by resumming the ladder diagrams
in the perturbation series, factoring it out and absorbing it in
the definition of the parton densities. In the case of the effective
theories we are concerned with, it must be remembered however that
(i) these theories are not asymptotically free, and (ii) the region
of interest to which one would like to use these is not the
asymptotic but the moderate $Q^2$ region.
\par All one can do here is expand the cross section in terms of
the effective quark-pion coupling constant $\alpha_\pi$.
\begin{equation}
\sigma(z,Q^2)=\delta(z-1)+{\alpha_\pi\over 2\pi}\sigma_\alpha
(z,Q^2)+O(\alpha_\pi^2)
\end{equation}
and redefine the parton densities as
\begin{displaymath}
q_i(x,Q^2)=q_{i\circ}(x)+\Delta q_i(x,Q^2)
\end{displaymath}
with
\begin{equation}
\Delta q_i(x,Q^2)={\alpha_\pi\over 2\pi}\sum_j\int^1_x
{dy\over y}q_{j\circ}(y)\sigma^{ij}_\alpha(x/y,Q^2)+O(\alpha^2_\pi)
\end{equation}
so that one recovers
\begin{equation}
F^{eN}_2(x,Q^2)/x=\sum_{i} e_{i}^2q_{i}(x,Q^2)
\end{equation}
which is the key formula in any theory of this kind. In $(2.4),~ \sigma^{ij}_\alpha$
denotes the lowest order cross section for the process$\gamma^*q_j
\rightarrow q_i\pi$.
\par Then with the standard notation $\tau=log(Q^2/\mu^2)$ 
, $\mu^2$ being a parameter defining the $Q^2$ scale one can
write down the evolution equation to the lowest order in $\alpha_\pi$
\begin{equation}
{d\over{d\tau}}q_i(x,\tau)=\sum_{j}{\cal P}_{ij}\otimes q_j 
\end{equation} 
where 
\begin{equation}
{\cal P}_{ij}\otimes q_j=\int{dy\over y}{\cal P}_{ij}(x/y,\tau)
q_j(y,\tau)
\end{equation}
\begin{equation}
{\cal P}_{ij}(z,\tau)={d\over {d\tau}}\sigma^{ij}_\alpha(z,\tau)
\end{equation}
The last equation does not by itself guarantee universality
or process-independence of the evolution kernels, an elegant
feature of the splitting functions appearing in the leading log
QCD evolution.

The crucial point is that emission or absorption of an isovector
causes a change of flavor of the initial quark.The lowest order cross
section for the process $\gamma^{*} q\rightarrow q^{\prime}\pi$ can be
written as (after taking out the scale factor)
\begin{equation}
\sigma _{\alpha}^{q\prime q}(x,Q^2)=\sigma _{d}^{q\prime q}(x,Q^2)
+\frac {e_{q\prime}}{e_{q}}\sigma _{i}^{q\prime q}(x,Q^2)
\end{equation}
The cross-section for the process $\gamma^*\pi\rightarrow q^\prime
\bar q$ can be written as 
\begin{equation}
\sigma_\alpha^{q^\prime \pi}=\sigma_d^{q^\prime \pi}(x,Q^2)+\left(
\frac{e_\pi}{e_q^\prime}-1\right)\sigma^{q^\prime\pi}_i(x,Q^2)
\end{equation}
where $\sigma _d$ is the direct ($t$-channel exchange) cross section and 
$\sigma _i$ arises from the interference with the crossed channel 
($s$-channel exchange) diagram. 
Both the terms are of the same order in $\alpha _\pi$ and none 
of them can be neglected in the kinematic region of moderate $Q^2$ 
and moderate energy to which the theory is intended to be applied.

On the other hand, the cross section for the process $\gamma^*q
\rightarrow \pi q^\prime$ can be written as
\begin{equation}
\sigma_\alpha^{\pi q}(x,Q^2)=\sigma_d^{\pi q}(x,Q^2)
\end{equation}
This together with (2.8) leads to evolution kernels such that
\begin{equation} 
{\cal P}_{q^\prime q}(1-x,Q^2)\not= {\cal P}_{\pi q}(x,Q^2)
\end{equation}
This makes it impossible to sustain the probabilistic meaning of the
evolution kernels. Further, the probabilistic interpretation would contradict
the flavor-dependence of the evolution kernels 
${\cal P}_{u\pi^+}\not={\cal P}_{d\pi^-}$ 
as that would apparently violate the SU(2) symmetry. While this adds
to the flavor-asymmetry of the sea distribution [9] which in any case
would arise in such a theory due to a multiplicatively evolving part
of the Gottfried sum [2] and may help explain the observed asymmetry,
the probabilistic interpretation of the evolutions is clearly traded off.

\section{Evolution equation in chiral field theory}
For a specific case, we focus our attention to the chiral field
theory [7] which was invoked by Eichten et al.[4] in their explanation
of the NMC result on the Gottfried sum rule. The linearized effective
lagrangian is
\begin{equation} 
{\cal L}_{\Pi q}=-\frac{g_A}{f}{\bar \psi}{\partial}_\mu\Pi\gamma^\mu
\gamma^5\psi
\end{equation}
Here $\Pi$ is the pseudoscalar octet of flavor SU(3) playing the role
of Goldstone bosons, $\psi$ the quark field, $g_A$ the dimensionless
axial vector coupling constant and $f$ the pseudoscalar decay constant.\\
\\
{\it Contributions from direct cross sections}\\
\\
1. $\sigma_d(\gamma^*q\rightarrow \pi q^\prime)$\\
The cross section for the process $\gamma^*q\rightarrow \pi q^\prime$
arising from the direct(t channel) term is found to be of the form
\begin{equation}
\sigma_d(\gamma^*q\rightarrow \pi q^\prime)=\frac{g_{A}^2}{f^2}\frac{\sigma_\circ
}{64\pi^2}\frac{1}{P^2}{2}{{m_{\pi}}^2}(m_q+m_{q^\prime})^2
\int^{t_2}_{-\Lambda^2}dt\frac{(m_q-m_q^\prime)^2-t}
{(t-m_{\pi}^2)^2}
\end{equation}
where
\begin{eqnarray}
\Lambda^2&=&min[-2m_q^2+2\sqrt{(P^2+m_q^2)(K^2+m_q^2)}
           +2PK,\Lambda^2_{\chi SB}]\\
t_2&=&2m_q^2-2\sqrt{(P^2+m_q^2)(K^2+m_q^2)}+2PK\\
P^2&=&\frac{(s+Q^2+m_q^2)^2}{4s}-m_q^2\\
K^2&=&\frac{(s-m_q^2-m_\pi^2)^2-4m_q^2m_\pi^2}{4s}
\end{eqnarray}
and $\Lambda^2_{\chi SB}$ is the cut-off parameter of the chiral field theory 
and corresponds to the scale of chiral symmetry breaking. The flavor 
factor has been suppressed in this and all the formulae given below.
\\ 
2. $\sigma_d(\gamma^*q_i\rightarrow q_j\pi)$\\ 
The cross section for the process $\gamma^*q_i\rightarrow q_j\pi$
arising from the direct (t channel) term is found to be of the from
\begin{equation}
\sigma_d(\gamma^*q_i\rightarrow q_j\pi)=\frac{g_A^2}{f^2}\frac{\sigma_\circ
}{64\pi^2}\frac{1}{P^2}\int^{t_2}_{-\Lambda^2}dt\frac{a_1+b_1t+c_1t^2+d_1t^3}{(t-m_q^2)^2}
\end{equation}
where
\begin{eqnarray}
\Lambda^2&=&min[2\sqrt{(P^2+m_q^2)(K^2+m_\pi^2)}+2PK-(m_q^2+m_\pi^2),
            \Lambda_{\chi SB}]\\
t_2&=&-2\sqrt{(P^2+m_q^2)(K^2+m_\pi^2)}+2PK+m_q^2+m_\pi^2\\
K^2&=&\frac{(s-m_q^2-m_\pi^2)^2-4m_q^2m_\pi^2}{4s}\\
P^2&=&\frac{(s+Q^2-m_q^2)^2+4m_q^2Q^2}{4s}
\end{eqnarray}
The expressions for the coefficients $a_1,~b_1,~c_1$ and $d_1$ in the integrand
are given in the appendix.\\
\\
3. $\gamma^*\pi\rightarrow q_i\bar {q_j}$\\
The cross section for the process $\gamma^*\pi\rightarrow q_i\bar{q_j}$
arising from the direct (t channel) term is found to be of the from
\begin{equation}
\sigma_d(\gamma^*\pi\rightarrow q_i\bar q_j)=\frac{g_A^2}{f^2}\frac{\sigma_
\circ}{32\pi^2}\frac{1}{K^2}\int^{t_2}_{-\Lambda^2}dt\frac{a_2+b_2t+c_2t^2}
{(t-m_q^2)^2}
\end{equation}
where
\begin{eqnarray}
\Lambda^2&=&min[\frac{1}{2}(s+Q^2-m_\pi^2)+2PK-m_q^2,\Lambda^2_{\chi SB}]\\
t_2&=&-\frac{1}{2}(s+Q^2-m_\pi^2)+2PK+m_q^2\\
K^2&=&\frac{(s+Q^2+m_\pi^2)^2}{4s}-m_\pi^2\\
P^2&=&\frac{s-4m_q^2}{4}
\end{eqnarray}
The expressions for the $a_2$, $b_2$ and $c_2$ in the integrand are given in the
appendix.
\\
\noindent {\it Contributions from interference cross sections}\\
\\
1. $\sigma_i(\gamma^*q_i\rightarrow q_j\pi)$\\
The cross section for the process $\gamma^*q_i\rightarrow q_j\pi$
arising from the cross term is found to be of the from.
\begin{equation}
\sigma_i(\gamma^*q_i\rightarrow q_j\pi)=\frac{g_A^2}{f^2}\frac{1}{64\pi^2}
(\frac{e_{q_i}}{e_{q_j}})\frac{1}{P^2}\frac{1}{(s-m_q^2)}
\int^{t_2}_{-\Lambda^2}dt\frac{a_1^i+b^i_1t+c^i_1t^2}{(t-m_q^2)}
\end{equation}
The expressions for the coefficients $a^i_1$, $b^i_1$ and $c^i_1$ in the
integrand are given in the appendix.\\
\\
2. $\gamma^*\pi\rightarrow q_i\bar {q_j}$\\
The cross section for the process $\gamma^*\pi\rightarrow q_i\bar q_j$
arising from the crossed term is found to be of the form.
\begin{equation}
\sigma_i(\gamma^*\pi\rightarrow q_i\bar q_j)=\frac{g_A^2}{f^2}\frac{\sigma_
\circ}{32\pi^2}(1-\frac{e_\pi}{e_q})\frac{1}{K^2}\int^{t_2}_{-\Lambda^2}
dt\frac{a^i_2+b^i_2t+c^i_2t^2}{(t-m_q^2)(t+a^i_\circ)}
\end{equation}
where
\begin{equation}
a^i_\circ=Q^2/z-(m_q^2+m_\pi^2)
\end{equation}
The expressions for the coefficients $a^i_2$, $b^i_2$ and $c^i_2$ in the
integrand are given in the appendix.
\par The cross section for $\gamma^*q\rightarrow \pi q^\prime$
evaluated perturbatively in the lowest order leads to the evolution
kernel ${\cal P}_{\pi q}$ which has a $z$ dependence 
\begin{eqnarray}
{\cal P}_{\pi q}(z)&=&\frac {\alpha_\pi}{8\pi^2}{\sigma_\circ}m_q^2m_{\pi}^2Q^2
\left[\frac {dp^2}{dQ^2}\frac {1}{p^4}\int^{t_2}_{t_1}dt\frac {t}{(t-m_\pi^2)^2}
-\frac {1}{p^2}\frac {t_2}{(t_2-m_{\pi}^2)^2}\frac{dt_2}{dQ^2}\right]\nonumber\\
\end{eqnarray} 
where we have written $\alpha_\pi=g_A^2/f^2$.
Other quantities are defined as follows.
\begin{equation}
t_{2}=2m_q^2-\sqrt{(P^2+m_q^2)(K^2+m_{\pi}^2)}+2PK
\end{equation}
\begin{equation}
t_1=2m_q^2-\sqrt{(P^2+m_q^2)(K^2+m_{\pi}^2)}-2PK
\end{equation}
where $P$ and $K$ are the magnitudes of the three momentum of the initial hadron
and pion. In the above $\Lambda_{\chi SB}$  is an ultraviolet cut off parameter [4] which may be supposed to 
define the scale of chiral symmetry breaking.  
The evolution kernel ${\cal P}_{q^\prime q}(z) $ corresponding to the process
$\gamma^*q\rightarrow q^\prime \pi$ is
\begin{equation}
{\cal P}_{q^\prime q}(z)=g^q_d(z,Q^2)+\frac{e_q}{e_q^\prime}f^q_i(z,Q^2)
\end{equation}
and the evolution kernel ${\cal P}_{q^\prime \pi}(z)$ for the process
$\gamma^*\pi\rightarrow q^\prime\bar q$ is
\begin{equation}
{\cal P}_{q^\prime \pi}(z)=g^\pi_d(z,Q^2)+\left({e_\pi\over e_q^\prime}
-1\right)f^\pi_i(z,Q^2)
\end{equation}
The expressions for $g^q_d,~g^\pi_d,~f^q_i$ and $f^\pi_i$ are given in
the Appendix.
The quark densities are then found to satisfy the following equations
\begin{equation}
{du(z,Q^2)\over {d\tau}}={\cal P}_{uu}\otimes u+{\cal P}_{ud}\otimes d
+{\cal P}_{u\pi^\circ}\otimes \pi^\circ+{\cal P}_{u\pi^+}\otimes \pi^+
\end{equation}
\begin{equation}
{{d\bar u(z,Q^2)}\over {d\tau}}={\cal P}_{\bar u\bar u}\otimes \bar u+
{\cal P}_{\bar u\bar d}\otimes \bar d+{\cal P}_{\bar u\pi^\circ}\otimes
\pi^\circ+{\cal P}_{\bar u\pi^-}\otimes \pi^-
\end{equation}
\\
\pagebreak
{and the pion densities satisfy
\begin{equation} 
\frac{d\pi^+(z,Q^2)}{d\tau}={\cal P}_{\pi^+u}\otimes u+{\cal P}_{\pi^+
\bar d}\otimes \bar d
\end{equation}
\begin{equation}
\frac{d\pi^-(z,Q^2)}{d\tau}={\cal P}_{\pi^-d}\otimes d+{\cal P}_{\pi^-\bar u}
\otimes \bar u
\end{equation}
\begin{equation}
\frac{d\pi^\circ(z,Q^2)}{d\tau}=\sum {\cal P}_{\pi^\circ q_i}\otimes q_i
\end{equation}
The evolution kernels are explicitly flavor-dependent as stated in eqn(2.12).
This implies that the Goldstone isotriplet has a non-vanishing contribution
to the evolution of the non-singlet combinations $q={u-d}$ and ${\bar q}=
{\bar u}
-{\bar d}$. This again is a consequence of inclusion of the interference term
in the kinematic region of interest and is a feature characteristic of 
evolution through flavor-changing interactions. The moment equations resulting
from these evolution equations will be discussed in section 5. The dependence
of the evolution kernels on $Q^2$ is shown in Figures 1-5. The value of 
$\lambda_{\chi SB}$ is chosen to be $1170 MeV$.}
\pagebreak
{\begin{figure}
\psfig{figure=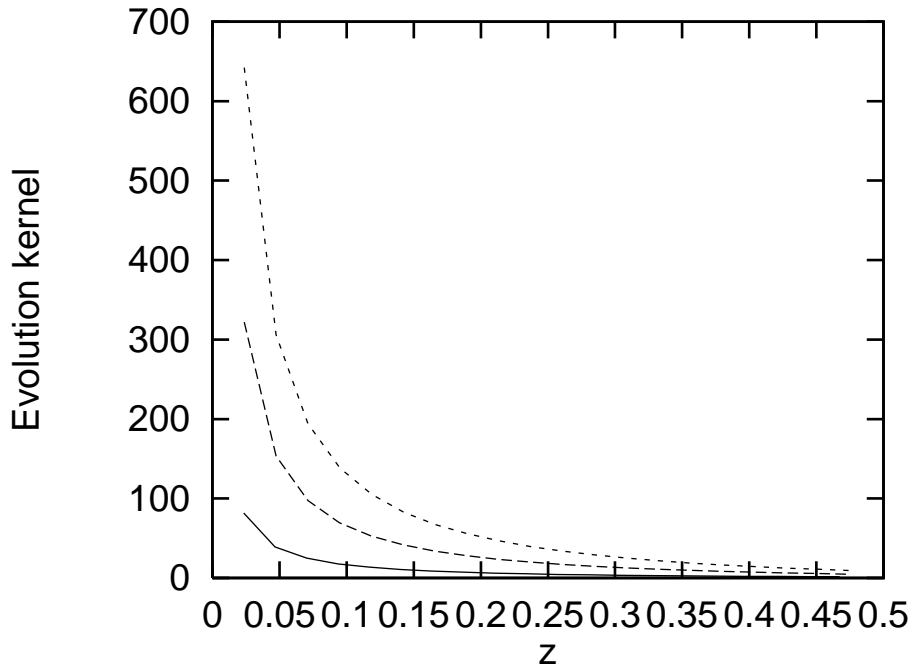,angle=-90}
\caption{Contribution of the t-channel diagram to $P_{q\pi}(z)$ for
1. $Q^2=2.0~ GeV^2$(solid line), 2. $Q^2=2.0~ GeV^2$(dashed line), and 3.
$Q^2=4.0~ GeV^2$(dotted line)}.   
\end{figure}
\pagebreak
\begin{figure}
\psfig{figure=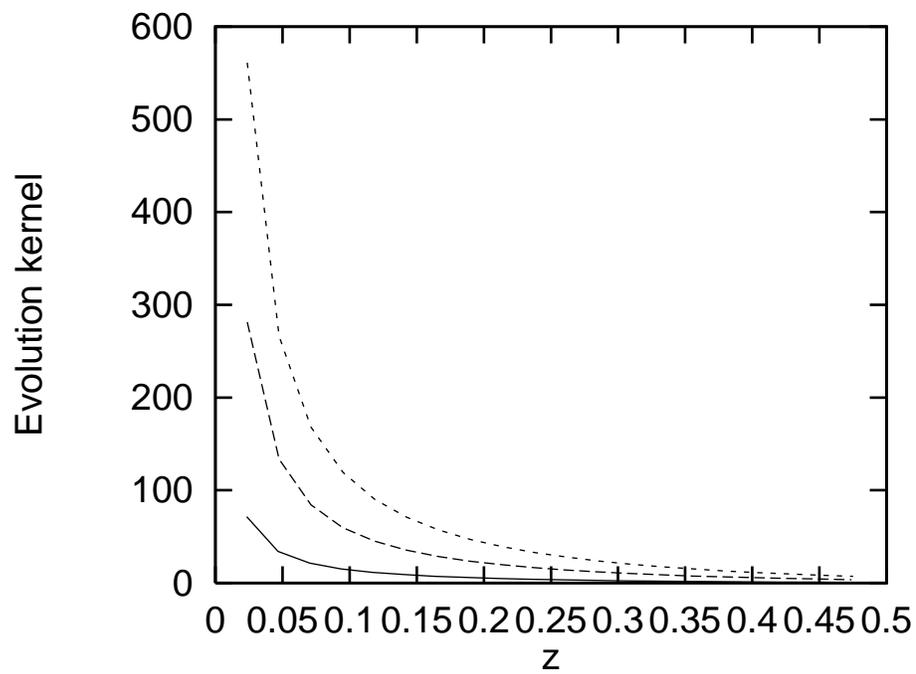,angle=-90}
\caption{Contribution of the interference term to the evolution kernel
$P_{q\pi}(z)$ for 1. $Q^2=0.5~ GeV^2$(solid line), 2. $Q^2=2.0~ GeV^2$(dashed
line), and 3. $Q^2=4.0~ GeV^2$(dotted line).}
\end{figure}
\begin{figure}
\psfig{figure=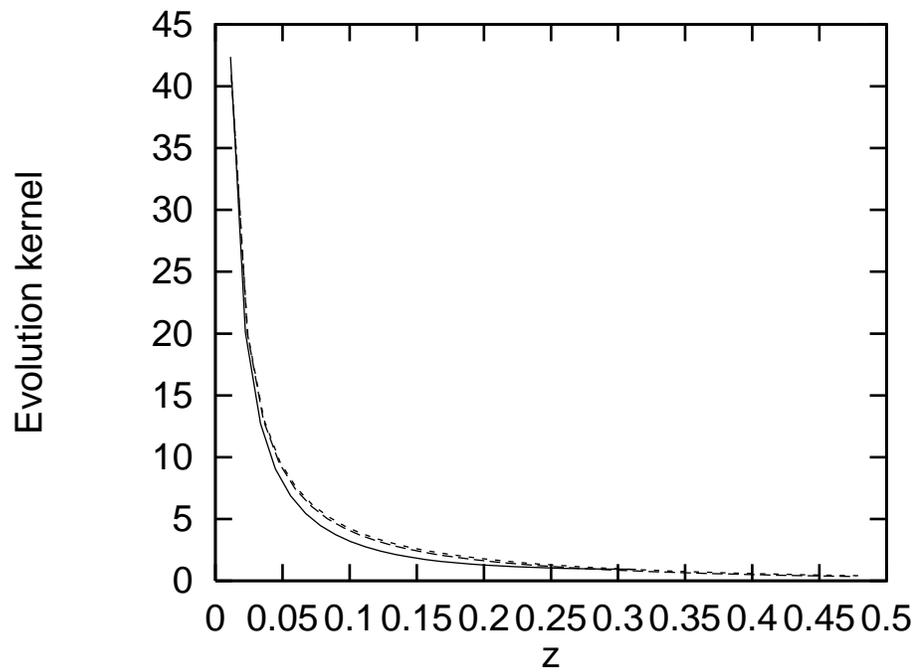,angle=-90}
\caption{Contribution of the t-channel diagram to the evolution kernel
$P_{ji}(z)$. The solid line corresponds to $Q^2=0.5~ GeV^2$ and the broken
line to $Q^2$ between 2.0 to 4.0 $GeV^2$.}   
\end{figure}
\begin{figure}
\psfig{figure=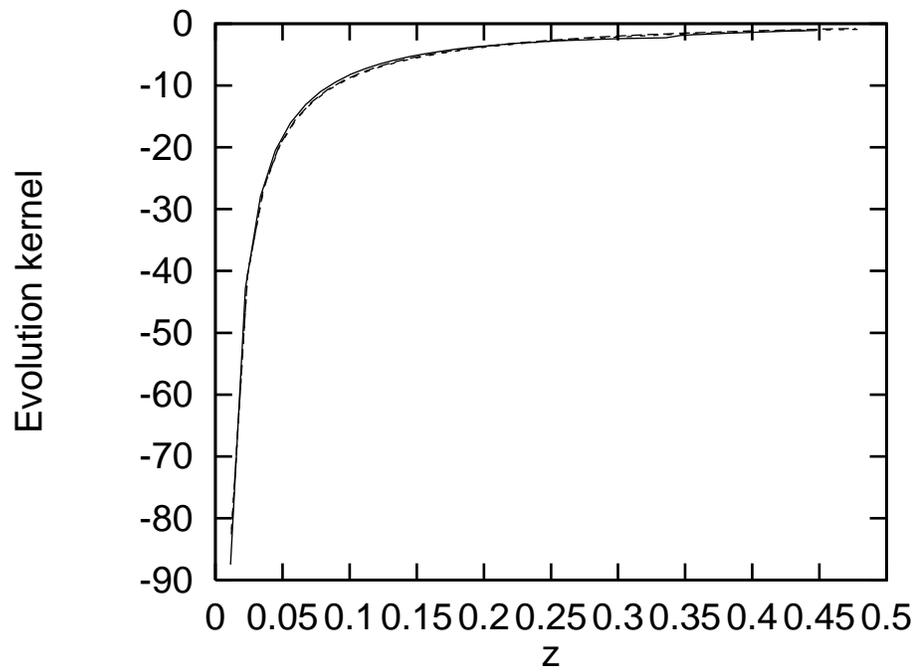,angle=-90}
\caption{Contribution of the interference term to the evolution kernel
$P_{ij}(z)$. The plot does not show any perceptible variation with $Q^2$
in the range $0.5-4.0~ GeV^2$.}
\end{figure}
\begin{figure}
\psfig{figure=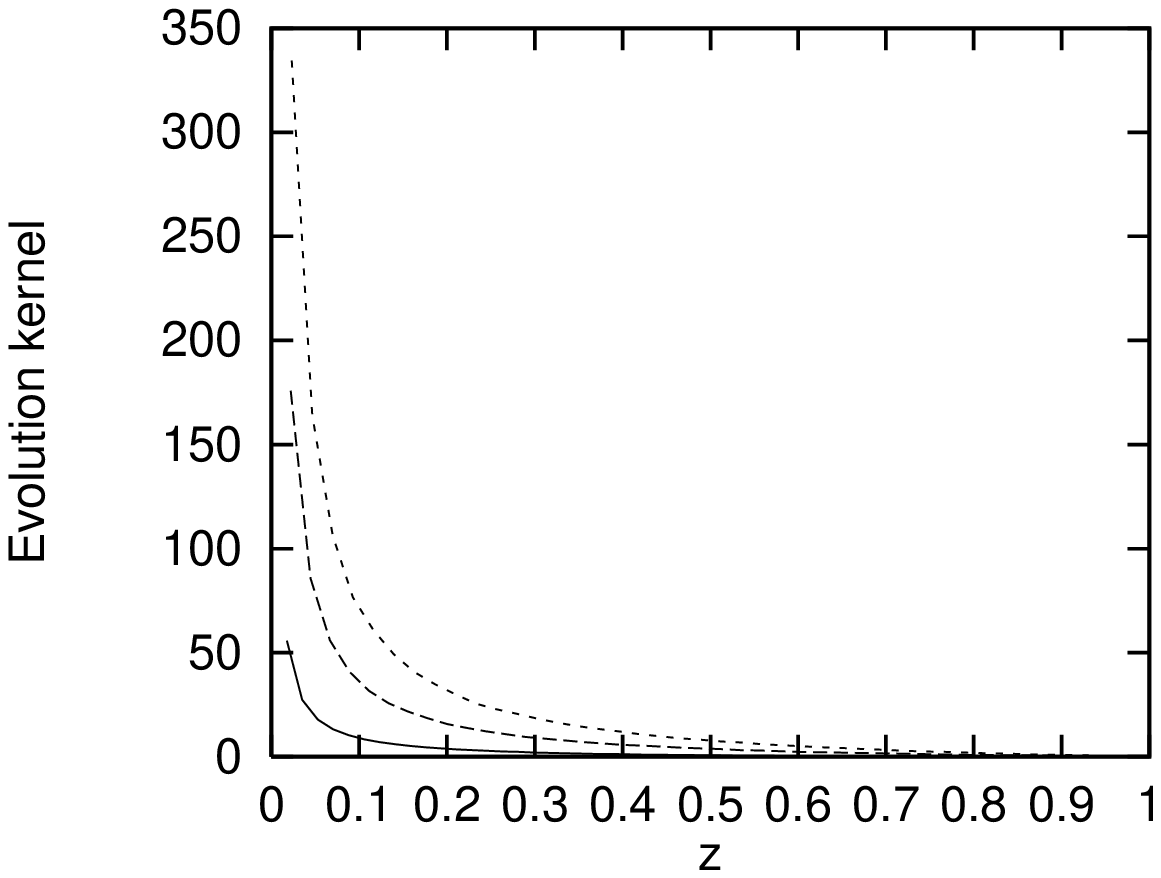,angle=-90}
\caption[]{Variation of the evolution kernel $P_{\pi q}(z)$ with $z$ plotted for
1. $Q^2=0.5~ GeV^2$(solid line), 2. $Q^2=2.0~ GeV^2$(dashed line), and 3. $Q^2
=4.0~ GeV^2$(dotted line).}
\end{figure}
\pagebreak}
{\section {Probabilistic calculation of the evolution kernels}
 In the kinematic region $xP>>Q\geq p_T$, where $P$ denotes the hadron
momentum, the leading logarithmic approximation is known to be valid
in QCD. The evolution kernels obtained in this approximation admits
probabilistic interpretation and can in fact be calculated taking
direct advantage of that. Durand and Putikka have shown that even the
infrared divergences can be eliminated in such calculations if the
`loss terms' are correctly incorporated. In this section, we apply
this method to derive the evolution equtions in the chiral field
theory for the kinematic region specified above. the exercise leads
to results which bring out striking differences between the evolution
kernels in QCD and those in 
$\chi FT$. In particular, it illustrates the 
difficulties one would face if one tries to apply the latter to high
energy.

Following Durand and Putikka, we now denote by $x$ the fraction
of the hadron momentum carried by a parton. 
Using the vertex factor of $\chi FT$, the transition probabilities
can be calculated from these diagrams in a straightforward
manner. Neglecting the quark masses and writing $\alpha_\pi=g_A^2/
f^2$, 
the resulting evolution equations are
\begin{eqnarray}
{\frac{du}{d\tau}}&=&\frac{\alpha_\pi m_\pi^2}{(2\pi)^3}\big[4xu(x)\int^x_
\circ
{dx_1{{x_1^3(x_1-x)}\over {(x_1-x)^2+\epsilon}}}\nonumber \\
& &+\int^1_x{dx_1\big[2x_1x^3q^+(x_1)+\frac{x_1-x}{2x_1}\pi^{(+\circ)}
(x_1)\big]\frac{x_1-x}{(x_1-x)^2+\epsilon}}\big]
\end{eqnarray}
\begin{equation}
\frac{d\pi^+}{d\tau}=\frac{\alpha_{\pi} m_{\pi}^2}{2x}{[\int_{x}^{1}
\frac{dx_1}{x_1}u(x_1)\frac{2(x_1-x)^3}{(x_1-x)^2 +\epsilon}
-\pi^{+}\int_{0}^{x}dx_1\frac{(x-x_1)^2}{(x-x_1)^2 +\epsilon}]}
\end{equation}
\begin{equation}
\frac{d\pi^0}{d\tau}=\frac{\alpha_\pi m_{\pi}^2}{x}{\left[\int_{x}^{1}
\frac{dx_1}{x}q^{+}(x_1)\frac{(x_1-x)^3}{(x_1-x)^2 +\epsilon}
-\pi^{0}\int_{0}^{x}dx_1\frac{(x-x_1)^2}{(x-x_1)^2 +\epsilon}\right]}
\end{equation}

where $\epsilon=p_T^2/2P^2$, $q^+=u+d$ and $\pi^{(+\circ)}=\pi^++\pi^\circ $.

There are obvious differences between these equations and those obtained
in QCD by the same procedure. The most important difference can be seen if
one considers the limit $\epsilon\rightarrow 0$. Whereas inclusion of
loss terms leads to elimination of infra-red divergences in the splitting
functions in the case of QCD evolution, the divergence in the $\epsilon
\rightarrow 0$ limit is not eliminated in the $\chi$FT evolution.
This shows that the kinematic region corresponding to this limit is beyond
the domain of applicability of the chiral field theory which is not valid
above the chiral symmetry breaking scale. This is true for all such theories
which would incorporate evolution through emission of chiral Goldstone bosons.

\section{Evolution of moments}
The $Q^2$-evolution of the moments of the structure functions in the 
chiral field theory follows directly from the basic equations (36) -
(40). However the resulting evolution is not as simple as in the case
of the leading logarithmic approximation. Complication arises due to two
distinctive features of the basic evolution equations. Firstly, the
kernels of evolution are $Q^2$-dependent. Secondly, the
flavor-dependence of the pionic contribution makes it harder to decouple
the moment equations. In particular, simple non-singlet combinations
like $q-\bar q$ no longer evolve in purely multiplicative manner.

We introduce the moments of the parton densities and the evolution
kernel by
\begin{equation}
Q_n(\tau )=\int_{0}^{1}dx~x^{n-1} q_n(x,\tau )
\end{equation}
\begin{equation}
\gamma {_n}^{AB}(\tau )=\int_{0}^{1}dxx^{n-1}{\cal P}_{AB}(x,\tau )
\end{equation}
and take moments of the equations (36) - (40) to obtain
\begin{eqnarray}
\frac {d}{d\tau}U_n(\tau )&=&\gamma{_n}^{uu}U_n +\gamma{_n}^{ud}D_n
+\gamma{_n}^{u\pi ^\circ }\Pi {_n}^{\circ}+\gamma{_n}^{u\pi^{+}}
{\Pi {_n}^{+}}\\
\frac {d}{d\tau }\bar U_n(\tau )&=&\gamma{_n}^{\bar u\bar u}\bar U_n 
+\gamma{_n}^{\bar u\bar d}\bar D_n +\gamma{_n}^{\bar u\pi^{\circ}}
\Pi _{n}^{\circ}+\gamma {_n}^{\bar u\pi^{-}}\Pi{_n}^{-}\\
\frac {d}{d\tau }\Pi{_n}^{+}(\tau )&=&\gamma {_n}^{\pi^{+}u}U_n 
+\gamma{_n}^{\pi ^{+}\bar d}\bar D_n\\
\frac {d}{d\tau }\Pi{_n}^{\circ }(\tau )&=&\sum \gamma{_n}^
{\pi ^{\circ}q_{i}}Q_{in}
\end{eqnarray}
and similar equations for $D_n$, $\bar D_n$, and $\Pi{_n}^{-}$.

We shall consider below the exact solution to these equations  
first and then make an approximation to bring out clearly the 
role of the pion decomposition process in the evolution of the 
moments. We shall omit the index $n$ from the symbols $Q_n, 
\gamma_n ^{qq}$ etc. in the expressions given below for reasons 
of typographical simplicity.
\subsection{\it Exact solutions}
Define the combinations $U^{(-)}=U-\bar U,~\Pi^{(-)}=\Pi^+ -\Pi^-$ etc. 
and transform to 
$$u^{(-)}=\exp\left(\int_{\tau _0}^{\tau}d\tau \gamma^{uu}\right)U^{(-)}.$$ 
Further writing 
\begin{displaymath}
X=\left(\begin{array}{c}u^{(-)}\\d^{(-)}\\ \Pi^{(-)}\end{array}\right)
\end{displaymath}
a formal solution is obtained as 
\begin{equation}
X(\tau)=\exp \left(\int _{\tau_0}^{\tau}M~d\tau\right)~X(\tau_0)
\end{equation}
where
\[ M=\left(\begin{array}{ccc}0&\tilde\gamma^{ud}&\gamma^{u\pi}\\
\tilde\gamma^{du}&0&\tilde\gamma^{d\pi}\\ \tilde\gamma^{\pi u}&
\tilde\gamma^{\pi d}&0\end{array}\right)\]  
The evolution takes a simple form for the eigenfunctions of the 
matrix $M$. Therefore the first task is to diagonalize it.
After diagonalization one finds that there are three eigenfunctions 
of evolution given by 
\begin{equation}
v_i(\tau)=u^{(-)}+\eta d^{(-)}+\zeta \Pi^{(-)}
\end{equation}
where 
$$\eta =\frac {\lambda_i \Gamma^{du}+\Gamma^{d\pi}\Gamma^{\pi u}}
{\lambda_{i}^2+\tilde \Gamma^{d\pi}\Gamma^{\pi d}}$$
$$\zeta =\frac{1}{\lambda_i}\left[\Gamma^{\pi u}-\Gamma^{\pi d}
\frac{\lambda_i\Gamma^{du}+\Gamma^{du}\Gamma^{\pi u}}
{\lambda_{i}^2 +\Gamma^{d\pi}\Gamma^{\pi d}}\right ]$$
$$\Gamma^{ij}=\int_{\tau_1}^{\tau} d\tau \gamma ^{ij}$$
The $\lambda_i$'s, being the three eigenvalues of $M$, are the 
roots of the cubic equation
\begin{equation}
\lambda^3-\alpha\lambda+\beta =0
\end{equation}
\begin{equation}
\alpha =\tilde \gamma^{u\pi}\gamma^{\pi u} -\tilde \gamma^{d\pi}
\gamma^{\pi d}-\tilde\gamma^{du}\tilde \gamma^{ud}
\end{equation}
\begin{equation}
\beta=\tilde \gamma^{du}\tilde \gamma ^{u\pi}\gamma^{\pi u}
-\tilde\gamma ^{ud}\gamma^{d\pi}\gamma^{\pi u}
\end{equation}
The evolution of each such combination $v_i (\tau)$ is then given
 by
$$v_i (\tau)=v_i (\tau_1)\exp (\lambda_i (\tau ,\tau_1)$$
It may be noted that the quantities $\Gamma^{ij}$ are simply
related to the moments of the cross section. Thus, putting 
back the moment index,
$$\Gamma_{n}^{ij}=\sigma_{n}^{ij}(\tau)-\sigma_{n}^{ij}(\tau_1)$$
\vskip 2cm
\subsection{\it {The role of the pion splitting function}}
Influnce of the pion splitting kernel on the evolution of the
non-singlet first moment can be clearly shown in a simplified
approximate version of the moment equations(5.81)-(5.84). 
Let us introduce the symbols ${\cal P}_{qq}^D={\cal P}_{q_iq_j},~ 
{\cal P}^{ND}_{qq}={\cal P}_{q_iq_j}~(i\neq j),~{\cal P}_{q\bar q}
^D={\cal P}^D_{q_i\bar q_i},~Q_{qq}={\cal P}^D_{qq}-{\cal P}^{ND}
_{qq} ,~Q_{q\bar q}={\cal P}^D_{q\bar q}-{\cal P}^{ND}_{q\bar q}$
where we assume the validity of the isospin SU(2) symmetry and charge
conjugation invariance.The moments of the kernels defined above will
be represented by $\gamma^D_n$ etc, in obvious analogy with(5.80).
Further, if we take the combination
$$q^+=(u-d)+(\bar u-\bar d)$$
and neglect the flavour-dependence of all other kernels except that
of ${\cal P}_{q\pi}$, the evolution equation for the combination $q^+$
of the quark and antiquark densities takes the simple form.
\begin{equation}
\frac{d}{d\tau}q^+(z,Q^2)=Q_{qq}\otimes q^+ +{\cal P}^\prime_\pi
\otimes (\pi^+-\pi^-)
\end{equation}
where ${\cal P}^\prime_{\pi}={\cal P}_{u\pi^+}-{\cal P}_{d\pi^+}$.
Now $Q_{q\bar q}$ may be neglected as it involves a process which is
second order in $\alpha_\pi$. The equation for the first moment is
then given by
\begin{equation}
\frac{d}{d\tau}Q^+_1=(\gamma^D_{qq}-\gamma^{ND}_{qq})Q^+_1
+\gamma^\prime_{\pi}\Pi_1
\end{equation}
This equation can be solved to yield
\begin{equation}
Q^+_1(\tau)=\Delta(\tau-\tau_\circ)[Q^+_1(\tau_\circ)+\int^{\tau}_
{\tau_\circ}d{\tau^\prime}\gamma^{\prime}_{\pi}(\tau^\prime)\Pi_1
(\tau^\prime)\Delta(\tau^\prime,\tau_{\circ})]
\end{equation}
where the evolution factor is given by
\begin{equation}
\Delta(\tau,\tau_\circ)=exp\int^{\tau}_{\tau_\circ}[\gamma^D_{qq}
(\tau^\prime)-\gamma^{ND}_{qq}(\tau^\prime)]d{\tau^\prime}
\end{equation}
This also turns out to be 
\begin{equation}
\Delta(\tau,{\tau_\circ})=A~ exp[\frac{1}{2}\hat{\sigma_1}^{\pi^\circ}
(\tau)-\hat{\sigma_1}^{\pi^+}(\tau)]
\end{equation}
\begin{equation}
A=exp[\hat{\sigma_1}^{\pi^+}(\tau_\circ)-\frac{1}{2}\hat{\sigma_1}^{\pi^\circ}
(\tau_\circ)]
\end{equation}
In the low $Q^2$ region, the pion density may be supposed to be
small enough for one to neglect the second term in (55) in comparison
with the first term. This situation corresponds to a multiplicative
evolution of the Gottfried sum which evolves from the quark model value
to a lower value at moderate $Q^2$, as observed experimentally[2].
At higher $Q^2$, the evolution becomes more complicated due to the
presence of the second term. 
\section{Conclusion}
One can obtain $Q^2$-evolution of the moments of the structure functions
in the chiral field theory at low $Q^2$, and the Gottfried sum does show 
variation with $Q^2$. However, the kinematic region in which the theory
is applicable makes it difficult to compare the results of the theory
with the NMC and the other experimental results. Further, unlike the
splitting functions in the Altarelli-Parisi formalisim, the evolution 
kernels in such effective theories admit simple probabilistic interpretation.
    
\section{Appendix}
Given below are the coefficients appearing in the different cross sections in 
section 3.
{\it Coefficients of the cross section for the process 
$\gamma^*q_i\rightarrow q_j\pi$}\\
\begin{eqnarray}
a_1&=&m_q^4(s+Q^2)-4m_q^2m_\pi^2Q^2-m_q^4(4m_q^2-3m_\pi^2)\\
b_1&=&-2m_q^2(s+Q^2)+3m_q^2(3m_q^2+2m_\pi^2)\\
c_1&=&s+Q^2-6m_q^2-m_{\pi}^2\\
d_1&=&1
\end{eqnarray}
{\it Coefficients of the cross section for the process 
$\gamma^*\pi\rightarrow q_i\bar q_j$}\\
\begin{eqnarray}
a_2&=&m_q^4(s+2m_q^2)+4m_q^2m_\pi^2(Q^2-m_q^2)\\
b_2&=&-2m_q^2(s+2m_q^2)-4m_q^2m_\pi^2\\
c_2&=&s+2m_q^2
\end{eqnarray}
{\it Coefficients of the cross section for the crossed channel process 
$\gamma^*q_i\rightarrow
q_j\pi$}\\
\begin{eqnarray}
a^i_1&=&m_q^2(Q^2+2m_q^2+3m_\pi^2)(s-m_q^2-m_\pi^2)
        -m_q^2(m_q^2-m_\pi^2)(s+3m_q^2-m_\pi^2)\nonumber\\
     & &  +m_q^2(Q^2+m_q^2)(s+m_{\pi}^2-m_q^2)
         -2m_q^2(s(s-2m_q^2-m_{\pi}^2)+m_q^2(m_q^2-m_\pi^2))\nonumber\\
     & & -4m_q^2m_\pi^2(s-2m_q^2+m_\pi^2)\\
b^i_1&=&-2(s-m_q^2)(s+Q^2-m_q^2-m_\pi^2)+4m_q^2(2m_q^2+m_\pi^2)\\
c^i_1&=&-2(s+m_q^2)
\end{eqnarray}
{\it Coefficients of the cross section for the crossed channel process 
$\gamma^*\pi\rightarrow
q_i\bar q_j$}\\
\begin{eqnarray}
a^i_2&=&-m_q^2(Q^2+m_q^2)(s-2m_q^2)-m_q^2(Q^2+m_q^2)(s+2Q^2-2m_\pi^2)
        -2m_q^2(s+Q^2-m_q^2)^2\nonumber\\
     & &  +2m_q^2m_\pi^2(s+Q^2+m_q^2)
         -4m_q^2m_\pi^2(2m_\pi^2+3m_q^2)+8m_q^2m_\pi^2(s+Q^2)\\
b^i_2&=&-(s-2m_q^2)(2s+Q^2-2m_q^2-2m_\pi^2)-s(Q^2+m_q^2)
        -m_q^2(s+2Q^2-2m_\pi^2)\nonumber\\
     & &-2m_q^2(2s+2Q^2-2m_\pi^2-m_\pi^2)+8m_q^2m_\pi^2\\
c^i_2&=&-2s
\end{eqnarray}
\vfil\eject
{\noindent 
{\bf References}}\\
\begin{enumerate}
\item G.Altarelli, Phys. Rep.,{\bf 81} (1982).
\item P.Amaudruz {\it et al.}, Phys. Rev. Lett. {\bf 66} (1991) 2712;\\
M.Arnedo et al., Phy.Rev. {\bf D 50}(1994)R1;\\
M.R.Adams et al.(E 665 collaboration), Phys.Rev.Lett. {\bf75}(1995)1466;\\
Phys.Rev. {\bf D 54}(1996)3006;\\K.Ackerstaff(hermes collaboration),DESY-
HERMES-96-01(1996)
\item A.Baldit et al.(NA51 collaboration), Phys.Lett. {\bf B332}(1994)244.
\item For a comprehensive review, see S.Kumano, Phys.Rep. {\bf303}(1998),183.
\item A.W.Thomas, Nucl. Phys. {\bf A532} (1991) 271;\\ 
A.Signal, A.W.Schreiber and A.W.Thomas, Mod. Phys. Lett. {\bf A6} 
(1991) 271.
\item E.J.Eichten, I. Hinchliffe and C. Quigg, Phys. Rev. {\bf D45} 
(1992) 2269.
\item E.M.Henley and G.A.Miller, Phys.Lett. {\bf B251}(1990)453;\\
S.Kumano, Phys.Rev. {\bf D43}(1991)59;\\
S.Kumano and J.T.Londergan, Phys. Rev. {\bf D44} (1991) 271.
\item R.D.Ball and S.Forte, Nucl. Phys. {\bf B425} (1994) 516.
\item K.Roy-Maity and P.Dasgupta, Int. J. Mod. Phys. {\bf A13} 
(1998) 1785.
\item H.Georgi and A.V.Manohar, Nucl. Phys. {\bf B234} (1984) 189.
\end{enumerate}}

\end{document}